\documentclass[%
 aip,
 amsmath,amssymb,
 reprint,%
]{revtex4-1}

\usepackage{graphicx}% Include figure files
\usepackage{dcolumn}% Align table columns on decimal point
\usepackage{bm}% bold math
\usepackage[utf8]{inputenc}
\usepackage[T1]{fontenc}
\usepackage{mathptmx}
\usepackage{etoolbox}
\usepackage{float}
\usepackage{subfigure}
\usepackage{hyperref}
\usepackage{xcolor}
\hypersetup{
    colorlinks=true,
    linkcolor=cyan,
    filecolor=cyan,      
    urlcolor=cyan,
    citecolor=cyan,
}

\makeatletter
\def\@email#1#2{%
 \endgroup
 \patchcmd{\titleblock@produce}
  {\frontmatter@RRAPformat}
  {\frontmatter@RRAPformat{\produce@RRAP{*#1\href{mailto:#2}{#2}}}\frontmatter@RRAPformat}
  {}{}
}%
\makeatother
\begin{document}

\preprint{AIP/123-QED}

\title[An MPI-OpenMP mixing parallel open source FW-H code for aeroacoustics calculation]
{An MPI-OpenMP mixing parallel open source FW-H code for aeroacoustics calculation}
% Force line breaks with \\
\author{Keli Zhang}
%  \altaffiliation{School of Aeronautic Science and Engineering, Beihang University, Beijing, 100191, China}%Lines break automatically or can be forced with \\
\affiliation{ 
School of Aeronautic Science and Engineering, Beihang University, Beijing, 100191, China%\\This line break forced with \textbackslash\textbackslash
}%
\affiliation{%
LHD, Institute of Mechanics, Chinese Academy of Sciences, Beijing, 100190, China%\\This line break forced% with \\
}%

\author{Changping Yu}
%  \altaffiliation{LHD, Institute of Mechanics, Chinese Academy of Sciences, Beijing 100190, China}%Lines break automatically or can be forced with \\
\affiliation{%
LHD, Institute of Mechanics, Chinese Academy of Sciences, Beijing, 100190, China%\\This line break forced% with \\
}%

\author{Peiqing Liu$^{*,}$}%
% \email{lpq@buaa.edu.cn}
\affiliation{ 
School of Aeronautic Science and Engineering, Beihang University, Beijing, 100191, China%\\This line break forced with \textbackslash\textbackslash
}%

\author{Xinliang Li$^{*,}$}%
\email{lixl@imech.ac.cn}
\email{lpq@buaa.edu.cn}
\affiliation{%
LHD, Institute of Mechanics, Chinese Academy of Sciences, Beijing, 100190, China%\\This line break forced% with \\
}%
\date{\today}% It is always \today, today,

\begin{abstract}
  In this paper, a permeable surface nondimensional FW-H (Ffowcs Williams\text{-}Hawkings) acoustics analogy post-processing code with convective effect 
  and AoA (angle of attack) corrections, OpenCFD-FWH, has been developed. OpenCFD-FWH is now used as post processing code of our finite volume CFD solver
  OpenCFD-EC (Open Computational Fluid Dynamic code for Engineering Computation). However, OpenCFD-FWH can also be used by other CFD solvers with the
  specified data interface. 
  The convective effect is taken into account by using Garrick Triangle to switch the wind tunnel cases coordinate system to a moving model with fluid
  at rest coordinate system, which simplifies the FW-H integration formulation and improves the computational efficiency of the code. The AoA effect is
  also taken into account by coordinate transformation. 
  In order to validate the code, three cases have been implemented. The first two cases are a monopole and a dipole in a mean flow with AoA, and
  the results of the code and the analytical solution are practically identical. The third case is the well-known 30P30N configuration with a
  Reynolds number of 1.71$\times10^6$ and an AoA of $5.5^\circ$. OpenCFD-EC with IDDES (Improved Delayed Detached-eddy simulation) is utilized
  to obtain the flow field, and the result shows relative good agreement when compared to JAXA experiments. Moreover, the code is implemented
  in a hybrid parallel way with MPI and OpenMP to speed up computing processes (up to 538.5 times faster in the 30P30N validation case) and
  avoid an out-of-memory situation. The code is now freely available on \url{https://github.com/Z-K-L/OpenCFD-FWH}.
\end{abstract}

\maketitle

\section{\label{sec:level1}INTRODUCTION}
With the escalating demands for environmental protection, aeroacoustics noise has 
received considerable attention from both the industrial and academic sectors, 
especially in the aviation sector. Aircraft noise is restricting the development of 
airports. Hence, it is very important to conduct far-field noise evaluation during 
aircraft development and design stages. One of the most practical ways to evaluate far-field
noise of the aircraft is the hybrid CAA (computational aeroacoustics) approach, since the DNS 
(direct numerical simulation) for far-field noise in engineering problems is unrealistic.

The hybrid CAA method involves obtaining unsteady flow field through CFD solvers and then 
employing acoustic analogy equations to calculate far-field noise, which is widely adopted 
due to its substantial reduction in computational complexity. 
For example, Molina et al. \cite{molina_flow_2019} investigated tandem cylinder noise through
DDES (Delayed Detached-eddy simulation) and FW-H (Ffowcs Williams\text{-}Hawkings) acoustic analogy.
Ma et al. \cite{ma_aerodynamic_2020} investigated aeroacoustic characteristics of Swept Constant-Chord 
Half model with four different types of high-lift devices through IDDES 
(Improved Delayed Detached-eddy simulation) and FW-H acoustic analogy. Hu et al. \cite{hu_noise_2022} 
utilized implicit wall-resolved LES (Large Eddy Simulation) and FW-H acoustic analogy 
to explore the noise reduction mechanisms of TE (trailing edge) serrations. Chen et al. 
\cite{chen_numerical_2023} also used the hybrid method of LES and the FW-H acoustic analogy, to 
study the noise of flow across a cylinder with varying spanwise lengths. 
Souza et al. \cite{souza_dynamics_2019} carried out LBM (Lattice Boltzmann Method) simulation on the
30P30N high-lift configuration and applied FW-H acoustic analogy to compute the associated acoustic field.
Teruna et al. \cite{teruna_numerical_2021} analyzed the noise reduction effect of a fully resolved 
3-D printed porous TE utilizing LBM and FW-H acoustic analogy as well. DNS and FW-H acoustic analogy
were conducted by Turner and Kim \cite{turner_quadrupole_2022} to assess the importance of quadrupole 
noise in aerofoil flow separation or stall conditions. They acquired the quadrupole noise by calculating the 
relative difference between the FW-H results of the solid and permeable surface.

Currently, there are only a few open source codes available for FW-H acoustic analogy, such as libAcoustics developed 
by Epikhin et al. \cite{epikhin_development_2015} for OpenFOAM written in C++, SU2PY\textunderscore FWH developed 
by Zhou et al. \cite{zhou_aeroacoustic_2018} for SU2 written in python, and a Farassat 1A solver developed by 
Shen et al. \cite{shen_validation_2020} for HiFiLES written in C++. However, they all have some problems. First 
they do not support MPI (Message Passing Interface) parallel to accelerate the computing processes and reduce
memory usage by distributing the computing tasks across multiple nodes/computers, which is very important when facing
large datasets. Second, they only support FW-H integration solutions for solid surface, which do not account for
the quadrupole noise and unable to address cases with porous materials. Third, libAcoustics and SU2PY\textunderscore FWH 
require the installation of OpenFOAM and SU2 software, respectively. Fourth, libAcoustics and SU2PY\textunderscore FWH 
do not consider inflow with an AoA (angle of attack). Finally, they lack comprehensive tutorials, making it difficult for
others to use their codes with other CFD solvers.

Hence, the OpenCFD-FWH code has been developed for our compressible finite volume CFD solver OpenCFD-EC 
\cite{li2010direct} (Open Computational Fluid Dynamic code for Engineering Computation). More importantly, it can be utilized
by any other solvers with the right data structures. Alternatively, one can modify the data reading module of the code accordingly.
The code is based on a permeable surface FW-H integration solution with the Garrick Triangle \cite{garrick1953theoretical} applied
to simplify the equations. The inflow with an AoA is also taken into account. Additionally, the code is implemented in a hybrid
parallel way to accelerate the computing processes and reduce the memory requirement for a single node/computer.
The deployment of the code demands only an MPI library and a Fortran 90 compilation environment.
Furthermore, Matlab programs for monopole and dipole validation are provided to generate the required input data for tutorial purposes.

The rest of the paper is organized as follows. Sec.\ref{sec:level2} derived the permeable surface FW-H acoustic analogy methods with
convective effect and AoA correction. Sec.\ref{sec:level3} depicted the code structure and parallel implementation. Sec.\ref{sec:level4}
present the results of the code for three different validation cases. Finally, conclusions are given in Sec.\ref{sec:level5}.

\section{\label{sec:level2}FW-H Acoustic Analogy}
\subsection{\label{sec:level2A}Permeable Surface FW-H equation}
The FW-H equation \cite{ffowcs1969sound} for permeable surface has the form of:
\begin{equation}
  \begin{aligned}
      \Box^2c^2(\rho-\rho_0) &= \frac{\partial}{\partial t}\left[Q_n\delta(f)\right] \\
      &\quad -\frac{\partial}{\partial x_i}\left[L_i\delta(f)\right]+\frac{\partial}{\partial x_i\partial x_j}\left[T_{ij}H(f)\right],
  \end{aligned}
  \label{eq:FWH}
\end{equation}
where the $\square^2=1/c^2\partial^2/\partial t^2-\nabla^2$ is the D'Alembert operator, c is the sound speed, $\rho$ is the
density, $\rho_0$ is the density of the undisturbed medium, $\delta$ and $H$ are the Dirac delta and Heaviside function, respectively.
The moving surface is described by $f(\boldsymbol{x},t)=0$ such that $\hat{\boldsymbol{n}}=\triangledown f$ is the 
unit outward normal of the surface \cite{farassat2007derivation}. The $Q_n$, $L_i$ and Lighthill tensor stress $T_{ij}$ are defined as:
\begin{align}
  Q_n&=Q_i\hat{n}_i=\left[\rho_0v_i+\rho(u_i-v_i)\right]\hat{n}_i,\label{eq:Qn}\\  
  L_i&=L_{ij}\hat{n}_j=\left[P_{ij}+\rho u_i(u_j-v_j)\right]\hat{n}_j,\label{eq:Li}\\
  T_{ij}&=\rho u_iu_j+P_{ij}+c^2(\rho-\rho_0)\delta_{ij},\label{eq:Tij}
\end{align}
where $v_i$ is the component of the velocity of the moving surface, $u_i$ is the component of the velocity of the fluid, $\delta_{ij}$ is 
the Kronecker delta, and $P_{ij}$ is the stress tensor:
\begin{align}
  P_{ij}=(p-p_{0})\delta_{ij}-\sigma_{ij},\label{eq:Pij}
\end{align}
where $p_{0}$ is the ambient pressure, $\sigma_{ij}$ is the viscous stress tensor. Usually $\sigma_{ij}$ is a negligible source of sound 
and is neglected by almost any other FW-H implementations \cite{epikhin_development_2015,zhou_aeroacoustic_2018,shen_validation_2020,
farassat2007derivation}. Hence, $P_{ij}=(p-p_{0})\delta_{ij}$ is used in this paper.

\subsection{\label{sec:level2B}Integration solution for general cases}
Neglecting the quadrupole term in Eqs.~(\ref{eq:FWH}), and following the derivation procedure of Farassat 1A formulation \cite{farassat2007derivation}, 
the integral solution of the FW-H equation for permeable surface can be derived as:
\begin{align}
  4\pi p_{T}^{\prime}(\boldsymbol{x},t)& =\int_{f=0}\left[\frac{\dot{{Q}_i} \hat{n}_i+{Q_i} \dot{\hat{n_i}}}{r(1-M_r)^2}\right]_{ret}dS\nonumber\\
  &+\int_{f=0}\left[\frac{Q_n(r\dot{M_r}+c_0(M_r-M^2))}{r^2(1-M_r)^3}\right]_{ret}dS,\label{eq:pT}
\end{align}

\begin{align}
4\pi p_L^{\prime}(\boldsymbol{x},t)& =\frac1{c_0}\int_{f=0}\left[\frac{\dot{L}_r}{r(1-M_r)^2}\right]_{ret}dS\nonumber\\
&+\int_{f=0}\left[\frac{L_{r}-L_{M}}{r^{2}(1-M_{r})^{2}}\right]_{ret}dS\nonumber\\
&+\frac{1}{c_0}\int_{f=0}\left[\frac{L_r(r\dot M_r+c_0(M_r-M^2))}{r^2(1-M_r)^3}\right]_{ret}dS,\label{eq:pL}
\end{align}

\begin{align}
p'(\boldsymbol{x},t)=p'_T(\boldsymbol{x},t)+p'_L(\boldsymbol{x},t),\label{eq:p}
\end{align}
where $\boldsymbol{x}$ is the observer coordinate vector, t is the observer time, r is the distance between observer and source, 
$c_{0}$ is the sound speed of the undisturbed medium, the superscript "\textperiodcentered{}" means derivative over the source time $\tau$
, and the subscripts T and L represent the thickness and loading components, respectively. M is the Mach number vector of the moving surface 
with component $M_{i}=v_{i}/c_{0}$. $M_r$, $L_r$, and $L_M$ are defined as:
\begin{align}
  M_r&=M_i\hat{r}_i,\label{eq:Mr}\\
  L_r&=L_i\hat{r}_i,\label{eq:Lr}\\
  L_M&=L_iM_i,\label{eq:LM}
\end{align}
where $\hat{r}_i$ is the component of the unit radiation vector.

The subscript $ret$ in Eqs.~(\ref{eq:pT}), and (\ref{eq:pL}) means the 
quantities inside the square brackets are evaluated at retarded time:
\begin{align}
  \tau_{ret}=t-r_{ret}/c~.\label{eq:tau}
\end{align}

Despite the quadrupole term in Eqs.~(\ref{eq:FWH}) is omitted, the quadrupole source inside the permeable FW-H surface is still be 
accounted for by Eqs.~(\ref{eq:p}) according to Brentner and Farassat \cite{brentner1998analytical}.

\subsection{\label{sec:level2C}Integration solution for wind tunnel cases}
Eqs.~(\ref{eq:p}) is derived in a coordinate system that, the source is moving in a stationary medium with observers at rest 
in the far-field. In a wind tunnel case, where both the source and observer are stationary within a uniform flow with an AoA,
which is the common scenario in the majority of aircraft CFD cases, the Garrick Triangle \cite{garrick1953theoretical} can be
applied to transform the coordinate system. In this new coordinate system, the source is now moving in a stationary medium,
while observers remain stationary relative to the source. This will lead to a large simplification of the formulation and
increase the computational efficiency of the code.

First, let us assume that the mean flow has a velocity of $U_0$ along the positive $x_1$ axis direction. The retarded time of
Eqs.~(\ref{eq:tau}) will be changed to:
\begin{align}
  \tau_{ret}=t-R/c_0~.\label{eq:tau2}
\end{align}
where R is the effective acoustic distance between the source and the observer \cite{bres2010ffowcs}:
\begin{align}
  R&=\frac{-M_0d_1+R_*}{\beta^2},\label{eq:R}\\  
  R_*&=\sqrt{d_1^2+\beta^2[d_2^2+d_3^2]},\label{eq:R_s}\\
  \beta&=\sqrt{1-M_0^2},\label{eq:beta}\\
  &M_0=U_0/c_0,\label{eq:M0}
\end{align}
where $d_i=x_i-y_i$ is the component of distance between the observer and the source.

The component of the unit radiation vector is now altered to:
\begin{align}
  \hat{R}_1&=\frac{-M_0R_*+d_1}{\beta^2R},\label{eq:R1}\\  
  \hat{R}_2&=d_2/R,\label{eq:R2}\\
  \hat{R}_3&=d_3/R~.\label{eq:R3}
\end{align}

Next, consider a mean flow with an AoA in the x-y plane, and its velocity magnitude remains equal to $U_0$. By using the 2D plane 
coordinate transformation
\begin{align}
  d_1^{\prime}&=\;\;\;\, d_1cos(\mathrm{AoA})+d_2sin(\mathrm{AoA}),\label{eq:d1_prime}\\  
  d_2^{\prime}&=        -d_1sin(\mathrm{AoA})+d_2cos(\mathrm{AoA}),\label{eq:d2_prime}
\end{align}
and bring them into the $d_1$ and $d_2$ of the Eqs.~(\ref{eq:R}), and Eqs.~(\ref{eq:R_s}) yields:
\begin{align}
  &\quad R=\frac{-M_1d_1-M_2d_2+R_*}{\beta^2},\label{eq:R_New}\\  
  R_*&=\sqrt{(M_1d_1+M_2d_2)^2+\beta^2[d_1^2+d_2^2+d_3^2]},\label{eq:R_s_New}\\
  &\qquad \; M_1=M_0cos(\mathrm{AoA}),\label{eq:M1}\\
  &\qquad \; M_2=M_0sin(\mathrm{AoA})~.\label{eq:M2}
\end{align}

The component of the unit radiation vector is also changed:
\begin{align}
  {\hat{R}_1}^{\prime}&=\frac{-M_0R_*+d_1^{\prime}}{\beta^2R},\label{eq:R1_prime}\\  
  {\hat{R}_2}^{\prime}&=d_2^{\prime}/R,\label{eq:R2_prime}\\
  \hat{R}_3&=d_3/R~.\label{eq:R3_New}\\
  \hat{R}_1&={\hat{R}_1}^{\prime}cos(\mathrm{AoA})-{\hat{R}_2}^{\prime}sin(\mathrm{AoA}),\label{eq:R1_New}\\  
  \hat{R}_2&={\hat{R}_1}^{\prime}sin(\mathrm{AoA})+{\hat{R}_2}^{\prime}cos(\mathrm{AoA})~.\label{eq:R2_New}
\end{align}

Then, replacing all the $r$ in Eqs.~(\ref{eq:pT}) $\sim$ Eqs.~(\ref{eq:Lr}) by Eqs.~(\ref{eq:R_New}).
In addition, both the moving surface and fluid velocity need to subtract the mean flow velocity, since
the coordinate system has changed. Now, the distance $R$ is a constant and can be calculated in advance
for each observer, rather than in every sampling frame. Moreover, the source time derivative of $\hat{n_i}$, 
and $M_R$ will be zero because the surface is in uniform rectilinear motion. Therefore, the simplified 
version of Eqs.~(\ref{eq:pT}), and Eqs.~(\ref{eq:pL}) for wind tunnel cases take the following form:
\begin{align}
  4\pi p_{T}^{\prime}(\boldsymbol{x},t)& =\int_{f=0}\left[\frac{\dot{{Q}_i} \hat{n}_i}{R(1-M_R)^2}\right]_{ret}dS\nonumber\\
  &+\int_{f=0}\left[\frac{Q_nc_0(M_R-M^2)}{R^2(1-M_R)^3}\right]_{ret}dS,\label{eq:pT_WT}
\end{align}

\begin{align}
4\pi p_L^{\prime}(\boldsymbol{x},t)& =\frac1{c_0}\int_{f=0}\left[\frac{\dot{L}_R}{R(1-M_R)^2}\right]_{ret}dS\nonumber\\
&+\int_{f=0}\left[\frac{L_R-L_M}{R^{2}(1-M_R)^{2}}\right]_{ret}dS\nonumber\\
&+\int_{f=0}\left[\frac{L_R(M_R-M^2)}{R^2(1-M_R)^3}\right]_{ret}dS,\label{eq:pL_WT}
\end{align}
with
\begin{align}
  &Q_n=\left[-\rho_0U_{0i}+\rho{u_i}\right]\hat{n}_i,\label{eq:Qn_New}\\  
  L_i&=[P_{ij}+\rho (u_i-U_{0i})u_j]\hat{n}_j,\label{eq:Li_New}\\
  &\qquad \; M_R=M_i\hat{R}_i,\label{eq:MR}\\
  &\qquad \; L_R=L_i\hat{R}_i~.\label{eq:LR}
\end{align}

Notice that the quantities inside the square brackets in Eqs.~(\ref{eq:pT_WT}), and (\ref{eq:pL_WT}) are now evaluated 
at the retarded time calculated by Eqs.~(\ref{eq:tau2}). And the necessary inputs for far-field noise calculation 
from the CFD solver include the coordinate, unit outward normal, and area of the FW-H surface, along with the density, 
velocity, and pressure pulsation at each sampling frame.

\subsection{\label{sec:level2D}Nondimensionalization}
Since OpenCFD-EC utilizes dimensionless Navier-Stokes equations, OpenCFD-FWH is based on a nondimensional version of
Eqs.~(\ref{eq:pT_WT}), and (\ref{eq:pL_WT}) to avoid data conversion errors and computational expenditures.

The reference quantity used for the dimensionless transformation is the mean flow quantity, with the exception that
the pressure is nondimensionalized by $\rho_0U_0^2$, and the coordinate is nondimensionalized by the units of the mesh,
which yields:
\begin{align}
  \rho^*=&\rho/\rho_0,\;\,
  u_i^*=u_i/U_0,\;\,
  v_i^*=v_i/U_0,\;\,
  p^*=p/{\rho_0U_0^2},\\
  c_0^*=&c_0/U_0=1/M_0,\;\,
  x_i^*=x_i/L_{ref},\;\,
  y_i^*=y_i/L_{ref},\\
  dS^*&=dS/{L^2_{ref}},\;\,
  t^*=tU_0/L_{ref},\;\,
  {\tau}^*={\tau}U_0/L_{ref}~.
\end{align}

Then the dimensionless FW-H integration solution for wind tunnel cases can be obtained by replacing the variables in 
Eqs.~(\ref{eq:pT_WT}), and (\ref{eq:pL_WT}) to their corresponding nondimensional counterparts:
\begin{align}
  4\pi {p_{T}^{\prime}}^*(\boldsymbol{x}^*,t^*)& =\int_{f=0}\left[\frac{\dot{{Q}^*_i} \hat{n}_i}{R^*(1-M_R)^2}\right]_{ret}dS^*\nonumber\\
  &+\int_{f=0}\left[\frac{Q^*_n(M_R-M^2)}{M_0{R^*}^2(1-M_R)^3}\right]_{ret}dS^*,\label{eq:pT_WT_nd}
\end{align}

\begin{align}
4\pi {p_L^{\prime}}^*(\boldsymbol{x^*},t^*)& =\int_{f=0}\left[\frac{M_0{\dot{L}_R}^*}{R^*(1-M_R)^2}\right]_{ret}dS^*\nonumber\\
&+\int_{f=0}\left[\frac{L^*_R-L^*_M}{{R^*}^2(1-M_R)^{2}}\right]_{ret}dS^*\nonumber\\
&+\int_{f=0}\left[\frac{L^*_R(M_R-M^2)}{{R^*}^2(1-M_R)^3}\right]_{ret}dS^*.\label{eq:pL_WT_nd}
\end{align}
with
\begin{align}
  &Q^*_n=\left[-U^*_{0i}+{\rho}^*u^*_i\right]\hat{n}_i,\label{eq:Qn_New_nd}\\  
  L^*_i&=[P^*_{ij}+{\rho}^* (u^*_i-U^*_{0i})u^*_j]\hat{n}_j,\label{eq:Li_New_nd}\\
  U^*_{01}=cos&(\mathrm{AoA}),\;\,
  U^*_{02}=sin(\mathrm{AoA}),\;\,
  U^*_{03}=0~.\label{eq:U_0_nd}
\end{align}

Note that $f=0$, $\hat{R}_i$ and $\hat{n}_i$ remain unchanged whether the formulations are 
nondimensional or dimensional. Thus, the superscript "*" will not be necessary.

\section{\label{sec:level3}Implementation of OpenCFD-FWH}
\begin{figure*}
  \includegraphics[width=0.95\textwidth]{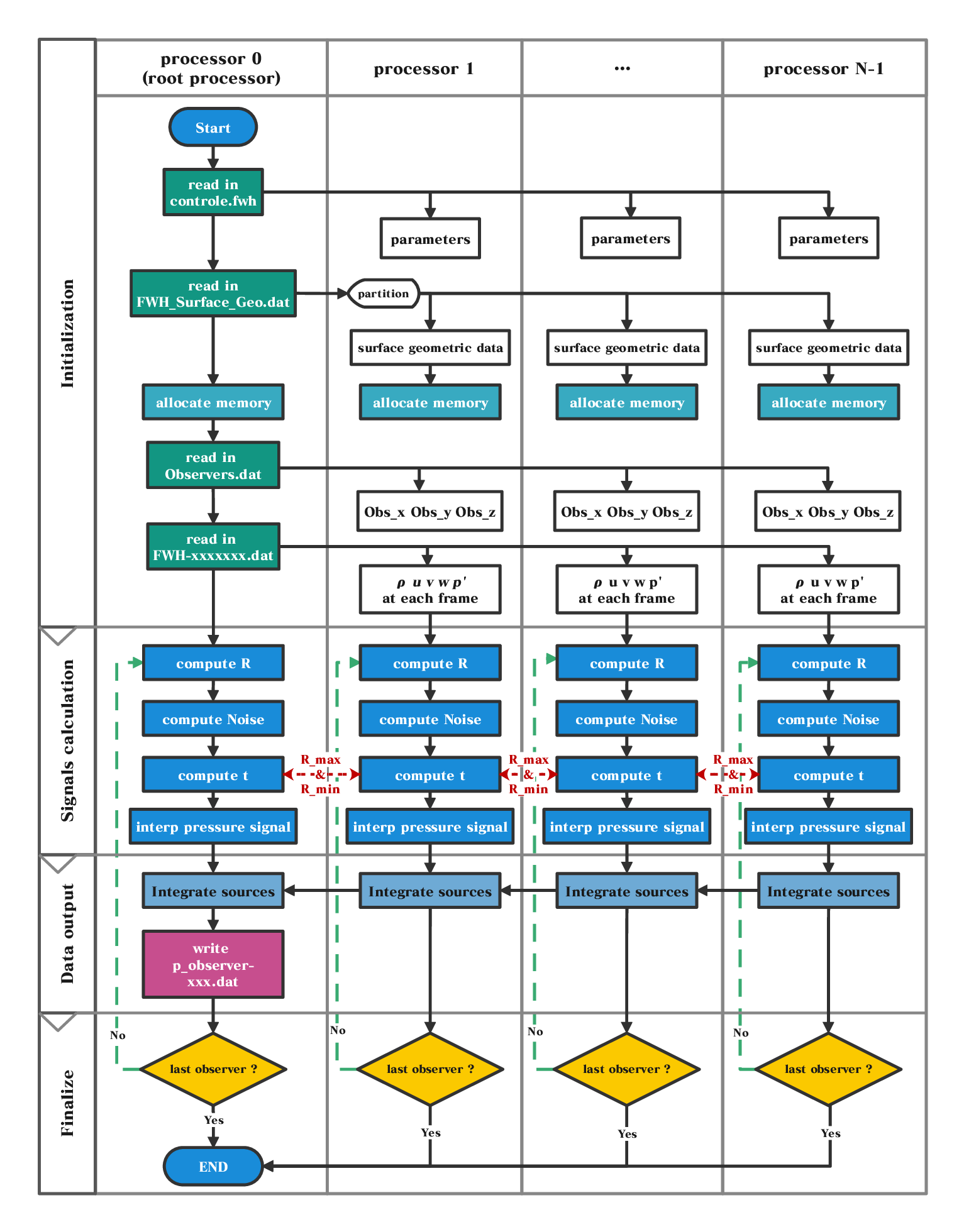}
  \caption{\label{fig:structure_of_the_code} MPI parallel framework of OpenCFD-FWH.}
\end{figure*}

OpenCFD-FWH can be divided into 4 main parts: initialization, pressure signals calculation, data output and finalization,
as illustrated in Fig.~\ref{fig:structure_of_the_code}.

\subsection{\label{sec:level3A}Initialization}
The first step of the code is to initialize all the MPI processors. This involves control file reading,
surface geometric data acquiring, assigning surface to the corresponding MPI processor, allocating memory,
and reading the location of the observers as well as the FW-H dataset.

The essential parameters for the code are specified in the control.fwh file, such as Mach number, AoA, time step,
number of observers, number of sampling frames, and number of OpenMP (Open Multi-Processing) threads. 
An example of the control.fwh file is given in Appendix~\ref{app:A1}.

The coordinate $y_i$, unit outward normal $\hat{n}_i$, and area $dS$ of each subface in the FW-H surface are included in
the FWH\textunderscore Surface\textunderscore Geo.dat file. These quantities are specified at the center of the subfaces 
and split in different Faces, due to OpenCFD-EC is a cell center solver for multiblock structure mesh. 
A detailed description of the file can be found in Appendix~\ref{app:A2}.

\begin{figure}
  \includegraphics[width=0.45\textwidth]{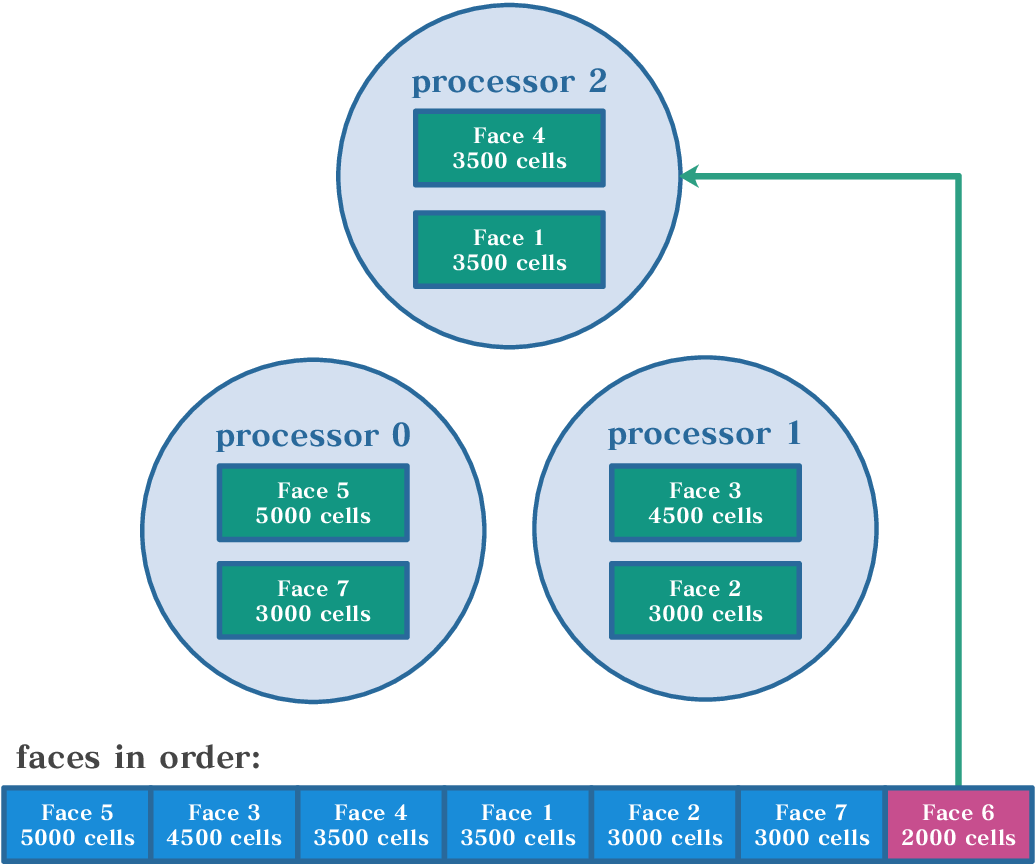}
  \caption{\label{fig:partition_method} Schematic of the MPI partition method for 3 processors and 7 Faces.}
\end{figure}
With the Faces information acquired, a partition method deployed by OpenCFD-EC for block splitting is applied for 
load balancing as shown in Fig.~\ref{fig:partition_method}. The method will first rank the Faces by their cell
numbers, and then assign the Faces to the processor with the least number of cells in order. As a result, the upper limit 
for the utilization of MPI processors by the code corresponds to the number of the Faces. To achieve a faster MPI 
acceleration result, one can divide FW-H surface as much and as equally as possible during the mesh generation stage,
or segment the output FW-H dataset of the CFD solver. Besides, when using a new big.little architecture 
CPU of Intel, or parallelizing an old system with a new one, one can adopt a partition method that considers the 
performance of the processors to accomplish optimal load balancing.

Following the MPI partition, each processor will allocate memory for FW-H data at all the sampling frames according to 
the assigned Faces. Then the root processor reads the observers.dat file, in which the coordinates of each observer occupy
a single row. Then, they are broadcasted to every other processor.

Subsequently, the FW-H dataset is read by the root processor and distributed to the corresponding processor.
An illustrative description of the FW-H dataset can be found in Appendix~\ref{app:A3}. Finally, 
memory is allocated for interpolated observer time across all processors, along with final pressure signal result
at the root processor.

\subsection{\label{sec:level3B}Pressure signals calculation}
The second step of the code is to calculate the pressure signals at one observer, rather than computing at all observers
at once for the sake of conserving memory usage. Additionally, OpenMP parallel is deployed on all processors to expedite
the calculation procedure. Further details regarding the MPI and OpenMP mixing parallel implementation will be expounded
in Section \ref{sec:level3D}.

The "compute R" subroutine in Fig.~\ref{fig:structure_of_the_code} is responsible for calculating the effective acoustic
distance based on Eqs.~(\ref{eq:R_New}) $\sim$ Eqs.~(\ref{eq:R2_New}).

The "compute Noise" subroutine in Fig.~\ref{fig:structure_of_the_code} is responsible for calculating the pressure signals
at each subface during respective source time, based on Eqs.~(\ref{eq:pT_WT_nd}) $\sim$ Eqs.~(\ref{eq:U_0_nd}). It is
noteworthy that $Q_n$ and $L_m$ remain unchanged in different observers, but they are both not stored to save memory.
In addition, $\dot{Q_n}$ and $\dot{L}_m$ are computed using second-order schemes, employing a one-sided scheme for the
first and last frame, while employing a central scheme for the other frames.

\begin{figure}
  \includegraphics[width=0.48\textwidth]{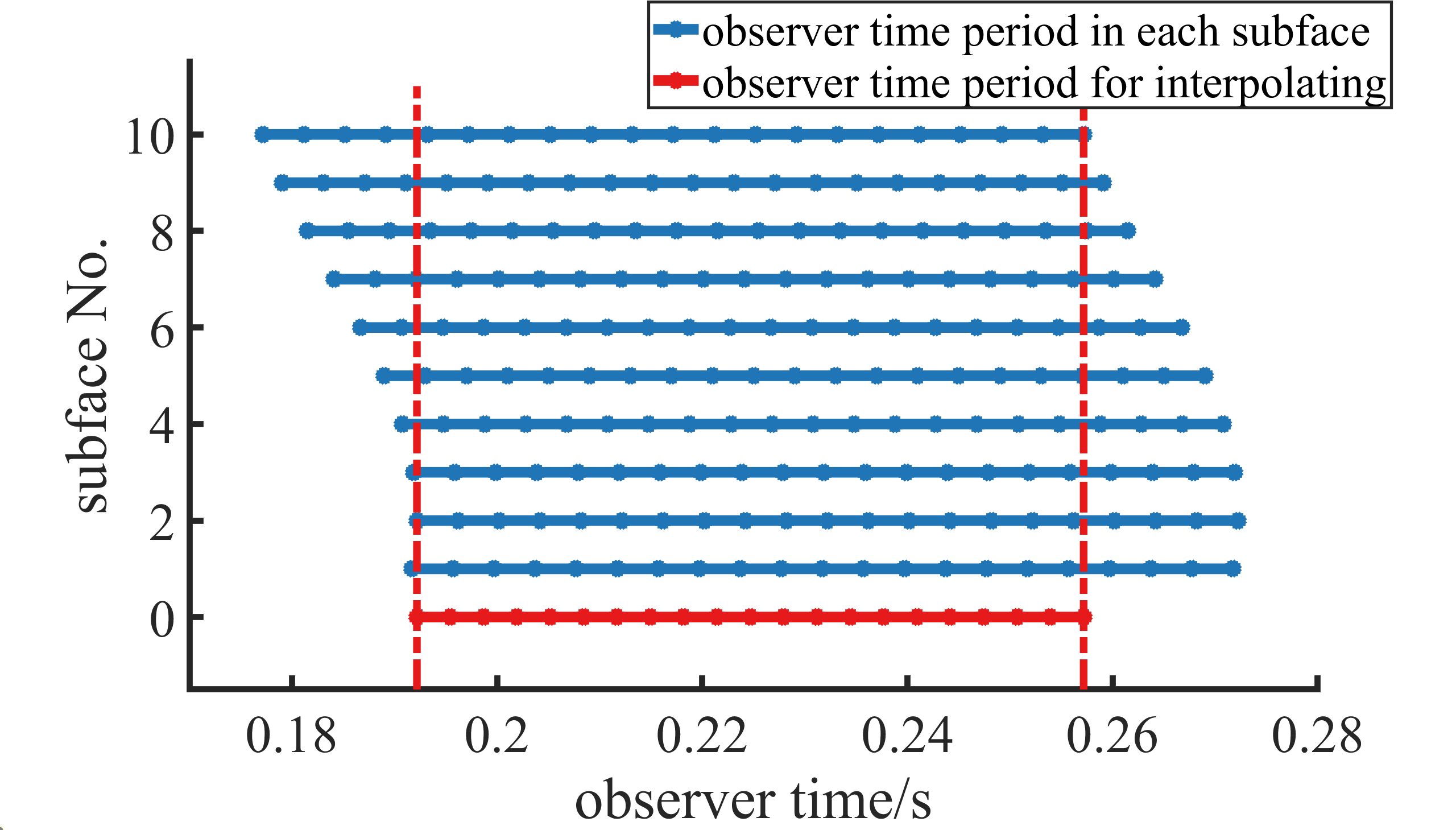}
  \caption{\label{fig:obs_time} A schematic representation of the observer time period regarding 10 different subfaces.}
\end{figure}

The "compute t" subroutine in Fig.~\ref{fig:structure_of_the_code} is responsible for calculating the observer time based on:
\begin{align}
  &t^*_{start}=R^*_{max}M_0,\label{eq:t_start}\\  
  t^*_{end}&=\tau_{max}+R^*_{min}M_0~,\label{eq:t_end}
\end{align}
where $R^*_{max}$ and $R^*_{min}$ are computed by making used of the MPI\textunderscore ALLREDUCE function. Consequently,
the observer time period, during which all subfaces collectively contribute to the observer pressure signal is
$t^*_{end}-t^*_{start}$, as shown in Fig.~\ref{fig:obs_time}.

The "interp pressure signal" subroutine in Fig.~\ref{fig:structure_of_the_code} is responsible for interpolating the
pressure signal of each subface depending on the source time, into pressure signal depending on the observer time, with
the help of cubic spline interpolation. After that, both the observer time and source time pressure signal stored at
every subface will be deallocated to conserve memory as well.

\subsection{\label{sec:level3C}Data output and Finalization}
The third step of the code is to conduct surface integration across all the subfaces to obtain the pressure signal at a
observer and output it in the p\textunderscore observer-xxx.log (xxx stands for the observer No.) file located within the
/FWH\textunderscore result-mpi folder.

The final step of the code is to verify whether all observers have completed their calculations, if the answer is no,
the code will loop back to the second step for the subsequent observers. If the answer is yes, all the processors will 
call MPI\textunderscore Finalize to end the code.
\subsection{\label{sec:level3D}Parallelization}
OpenMP is a widely used API (application programming interface) that supports shared-memory parallelization in multi-core 
and multi-processor systems. It is developed to facilitate parallel programming in C, C++, and Fortran, and can be easily 
deployed without extensive modifications to the existing serial codes.

OpenCFD-FWH is implemented in a hybrid parallel way that the FW-H data surface is spread to many MPI processors,
and OpenMP is deployed to split the loop in the computing stage among each MPI processor. This will result in an enormous
reduction of the computation time, as shown in Table~\ref{tab:parallelization}. With the use of 31 nodes, each with 32 CPU
cores for OpenMP parallelization, a remarkable acceleration of up to 538.5 times is achieved in comparison to the serial condition.
When the number of MPI processors (only MPI parallel) and OpenMP threads (only OpenMP parallel) is almost equal, their acceleration
effect is nearly the same.
\begin{table*}
\caption{\label{tab:parallelization} Initialization, computation, and overall execution time of OpenCFD-FWH
runs in different MPI processors and OpenMP threads for 18252 subfaces, 6535 sampling frames, and 40 observers of the 30P30N
validation case on the CAS SunRising platform. The platform has 32 cores x86 CPU on each node, with a based clock speed of 2.0GHz.
And the MPI environment deployed is the Intel MPI library.}
\begin{ruledtabular}
\renewcommand{\arraystretch}{1.25}
\begin{tabular}{ccccccc}
MPI processors&OMP threads&init time/s&computing time/s&total time/s&computing acceleration ratio&total acceleration ratio\\
\hline
 1 & $\times$ & 753.7 & 18120.3 &  18873.9 & $\backslash$ & $\backslash$ \\
31 & $\times$ & 887.2 &  811.3  &  1698.5  &     22.3     &     11.1     \\
 1 &    32    & 823.9 &  774.3  &  1598.1  &     23.4     &     11.8     \\
31 &    32    & 717.5 &  33.7   &  751.2   &     538.5    &     25.1     \\
\end{tabular}
\end{ruledtabular}
\end{table*}

Additionally, Table~\ref{tab:parallelization} illustrates that the predominant portion of the execution time is spent on
initialization in the hybrid parallel condition. This is attributed to the fact that I/O operations can not be accelerated,
as the data reading process requests sequential operations.

Furthermore, the 30P30N validation case costs a maximum of 62.6 GB of RAM (Random Access Memory). Without MPI parallelization, 
the computational demands for larger FW-H datasets can pose significant challenges for nodes and computers with limited memory
capacity. Hence, the MPI parallelization ensures successful execution on memory-constrained systems, or for even larger datasets
that can easily consume hundreds of RAM.

\section{\label{sec:level4}Validation}
Stationary monopole and dipole in a uniform flow with analytic solutions, along with a 30P30N case computed by
OpenCFD-EC solver are used to validate OpenCFD-FWH.
\subsection{\label{sec:level4A}Stationary monopole in a uniform flow with AoA}
The complex velocity potential for a stationary monopole in a uniform flow is given by Najafi et al.
\cite{najafi2011acoustic} as:
\begin{align}
  \phi(\boldsymbol{x},t)&=\frac{A}{4\pi R_*}\text{exp}\bigg[\mathrm{i}\omega\bigg(t-\frac{R}{c_0}\bigg)\bigg]~.\label{eq:phi_monopole}
\end{align}
where A is the amplitude, $\omega$ is the angular frequency of the monopole, and $i$ is the imaginary unit. In contrast to Najafi et al.
\cite{najafi2011acoustic}, Eqs.~(\ref{eq:R_New}) and (\ref{eq:R_s_New}) are used here to calculate $R$ and $R_*$, respectively,
taking into account the AoA effect of the uniform flow.

The velocity, pressure, and density pulsations induced by the monopole are:
\begin{align}
  &\qquad \quad \boldsymbol{u}^{\prime}(\boldsymbol{x},t)=\nabla\phi(\boldsymbol{x},t),\label{eq:u_monopole}\\
  p^{\prime}(\boldsymbol{x},t)&=-\rho_0\left[\frac\partial{\partial t}+U_{01}\frac\partial
  {\partial x_1}+U_{02}\frac\partial{\partial x_2}\right]\phi,\label{eq:p_monopole}\\
  &\qquad\qquad\rho^{\prime}(\boldsymbol{x},t)=\frac{p^{\prime}}{c_0^2},\label{eq:rou_monopole}\\
  U_{01}&=U_0cos(\mathrm{AoA}),\;U_{02}=sin(\mathrm{AoA})~.
\end{align}

The parameters used for the monopole are given in Table~\ref{tab:monopole}. To avoid any errors introduced by the CFD solver,
the FW-H dataset for the code is generated by Eqs.~(\ref{eq:phi_monopole}) $\sim$ (\ref{eq:rou_monopole}). 
\begin{table}[H]
\caption{\label{tab:monopole} Parameters for the monopole validation case.}
\begin{ruledtabular}
\renewcommand{\arraystretch}{1.25}
\begin{tabular}{cccccc}
$c_0\;\; m/s$&$M_0$&$\rho_0\;\; kg/m^3$&$AoA/{}^\circ$&$A\;\; m^2/s$&$f/Hz$\\
\hline
 340 & 0.6 & 1 & 45 & 1 & 5\\
\end{tabular}
\end{ruledtabular}
\end{table}

\begin{figure}
  \includegraphics[width=0.45\textwidth]{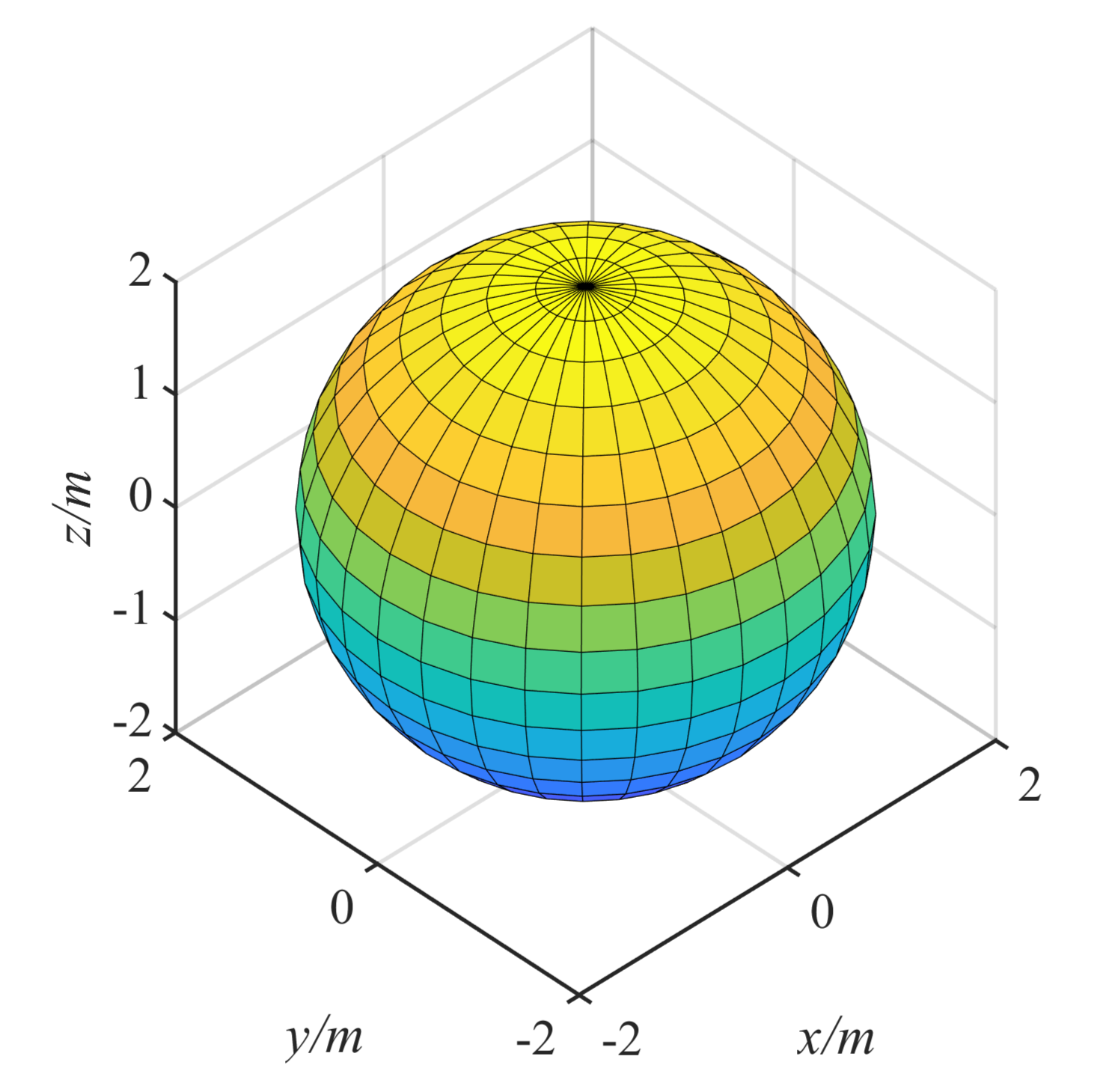}
  \caption{\label{fig:sphere} Schematic of the permeable FW-H data surface for monopole and dipole validation cases.}
\end{figure}

The permeable FW-H data surface is a sphere with a radius of 2 meters. Its center is located on the monopole. The sphere is
divided into 18 segments at the polar angle direction and 36 segments at the azimuth angle direction, resulting in a total of
648 subfaces, as shown in Fig.~\ref{fig:sphere}.

The observer locations are evenly distributed along a circle with a radius of 340 meters in the x-y plane, and its center is also
located on the monopole. A Matlab code is written to generate 1000 sampling frames, covering a time period of 2.5 seconds, in just
26 seconds on a laptop equipped with an Intel i7-13620H CPU. Subsequently, the OpenCFD-FWH code processes the dataset in less than
2.5 seconds with 12 OpenMP threads on only one MPI processor for 20 observers. Since the sphere FW-H surface is generated as a
single Face.

\begin{figure}
  \includegraphics[width=0.48\textwidth]{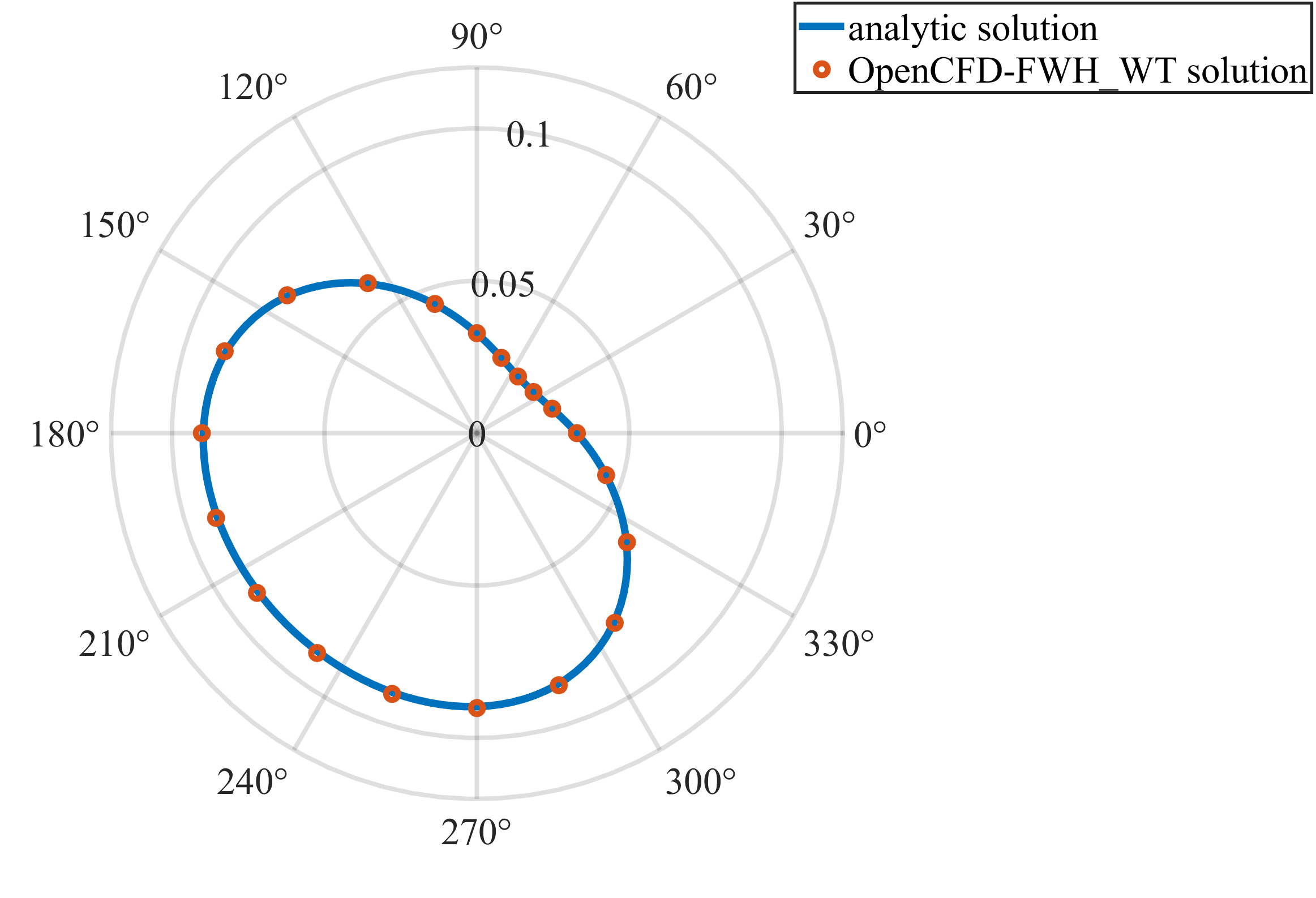}
  \caption{\label{fig:monopole_directivity} Far-field directivity of RMS pressure induced by the monopole at $r = 340\;m$.}
\end{figure}

\begin{figure}
  \includegraphics[width=0.47\textwidth]{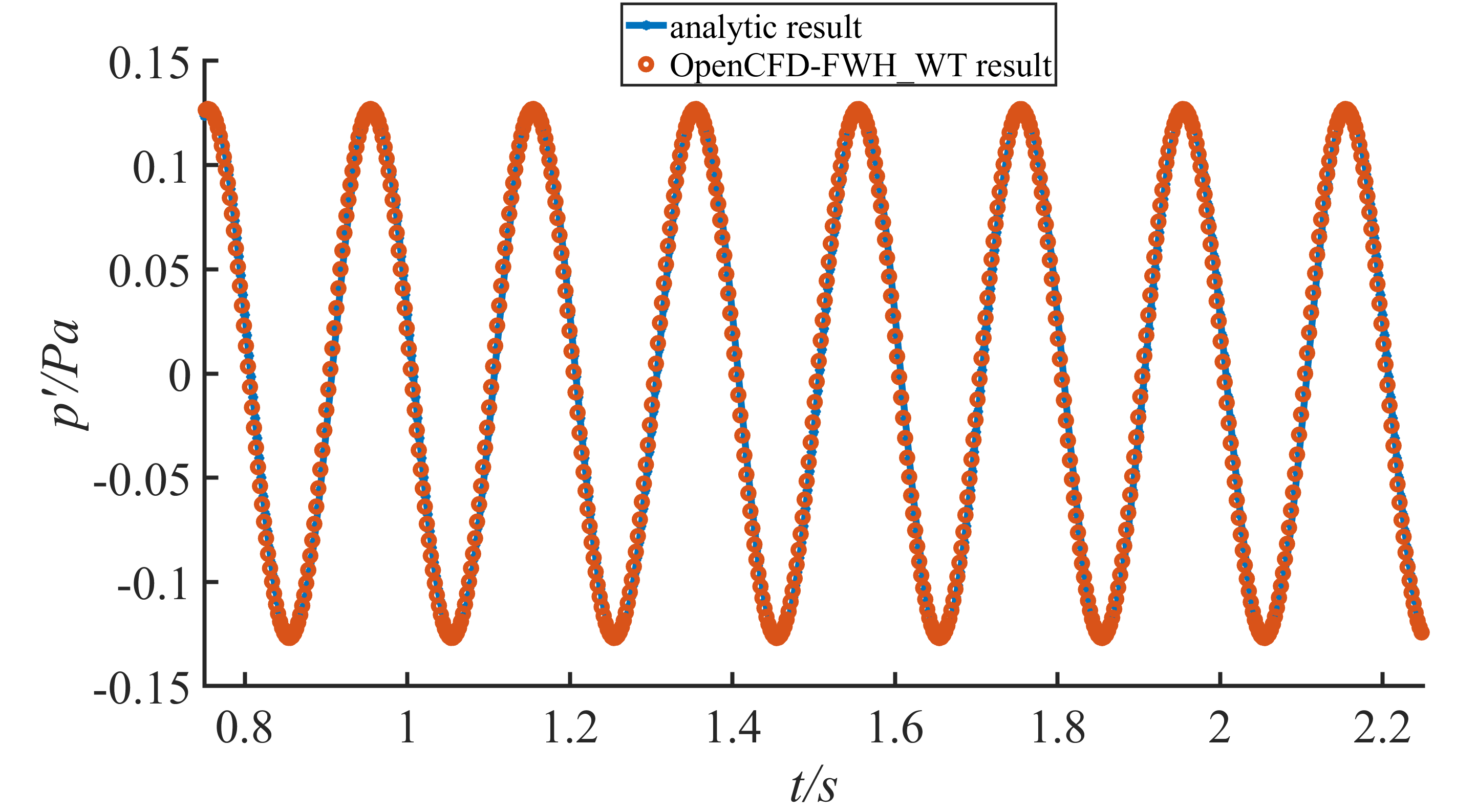}
  \caption{\label{fig:monopole_signal} Acoustic pressure signal at $(0\;m,-340\;m,0\;m)$ induced by the monopole}
\end{figure}

The comparison between the exact monopole solution and the result obtained from the OpenCFD-FWH code of far-field RMS
(root mean square) pressure directivity and pressure signal of the right below observer are shown in Fig.~
\ref{fig:monopole_directivity} and Fig.~\ref{fig:monopole_signal}, respectively. Very good agreements are observed between
the exact solution and the code. It is worth noting that results with even smaller errors can be achieved by using finer 
FW-H surface mesh and higher time sampling frequencies, but for the sake of simplicity, these results are not presented here.

Moreover, Fig.~\ref{fig:monopole_directivity} demonstrates that the directivity pattern of the monopole is diverted towards 
the inflow due to the convective effect.

\subsection{\label{sec:level4A}Stationary dipole in a uniform flow with AoA}
The complex velocity potential for a stationary dipole, with the polar axis coinciding with the x2-axis in a uniform flow
is given by:
\begin{align}
  \phi(\boldsymbol{x},t)&=\frac\partial{\partial x_2}\left\{\frac A{4\pi\textit{R}_*}\left.
    \exp\left[\mathrm{i}\omega\left(t-\frac R{c_0}\right)\right]\right\}\right.~.\label{eq:phi_dipole}
\end{align}

The velocity, pressure, and density pulsations induced by the dipole are acquired by Eqs.~(\ref{eq:u_monopole})
$\sim$ (\ref{eq:rou_monopole}) as well. And the parameters used for the dipole are given in Table~\ref{tab:dipole}.
\begin{table}[H]
\caption{\label{tab:dipole} Parameters for the dipole validation case.}
\begin{ruledtabular}
\renewcommand{\arraystretch}{1.25}
\begin{tabular}{cccccc}
$c_0\;\; m/s$&$M_0$&$\rho_0\;\; kg/m^3$&$AoA/{}^\circ$&$A\;\; m^2/s$&$f/Hz$\\
\hline
 340 & 0.5 & 1 & 10 & 1.5 & 7.5\\
\end{tabular}
\end{ruledtabular}
\end{table}

The FW-H data surface remains consistent with the monopole case, while the observer locations have been relocated
to a radius of 50 meters in the x-y plane. Another Matlab code has been developed to generate 1000 sampling frames, 
covering a time period of 2 seconds. The time required for generating the FW-H dataset and post-processing it is
basically the same compared with the monopole case.
\begin{figure}
  \includegraphics[width=0.47\textwidth]{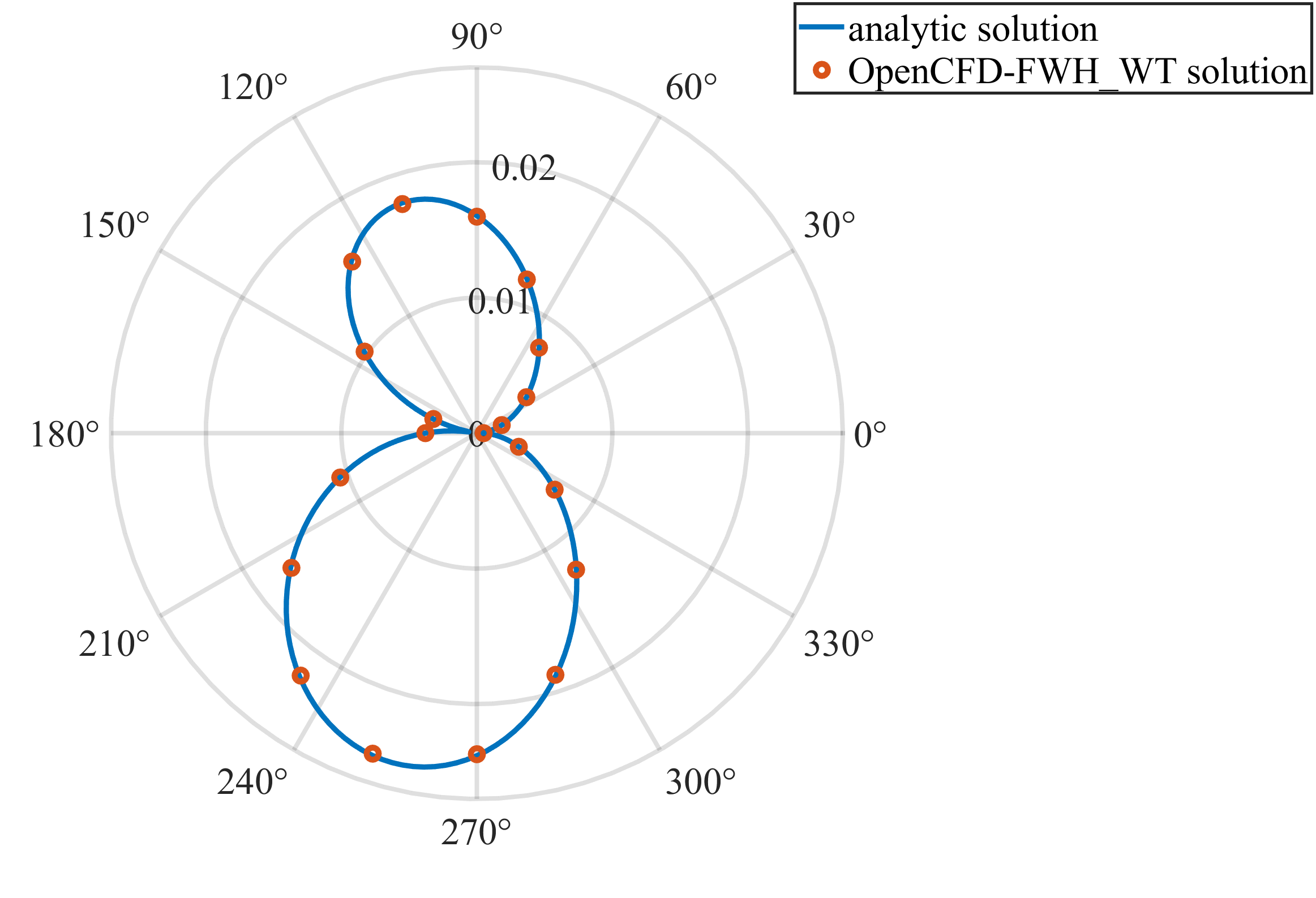}
  \caption{\label{fig:dipole_directivity} Far-field directivity of RMS pressure induced by the dipole at $r = 50\;m$.}
\end{figure}

\begin{figure}
  \includegraphics[width=0.48\textwidth]{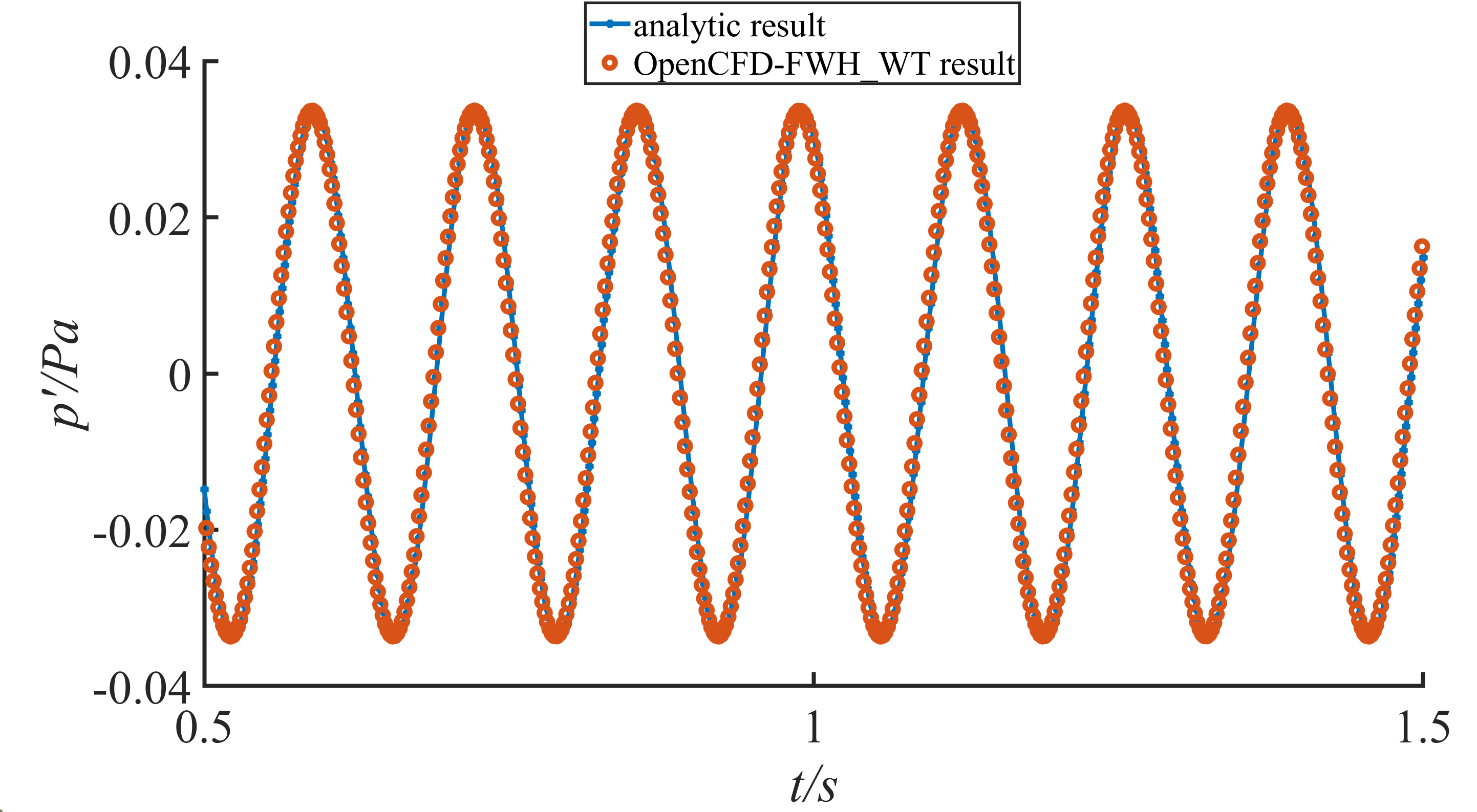}
  \caption{\label{fig:dipole_signal} Acoustic pressure signal at $(0\;m,-50\;m,0\;m)$ induced by the dipole.}
\end{figure}

Fig.~\ref{fig:dipole_directivity} and Fig.~\ref{fig:dipole_signal} present the result of the dipole far-field RMS pressure 
directivity and pressure signal of the right below observer, respectively. Excellent agreements are also achieved between 
the exact solution and the code. By applying finer FW-H surface mesh and higher time sampling frequencies, results with
essentially no error can be achieved. Again, for the sake of simplicity, these results are not presented here.

The mean flow leads to a reorientation of the maximum RMS pressure, resulting in a larger RMS pressure in the inflow
direction as shown in Fig.~\ref{fig:dipole_directivity}.

All the Matlab programs used for the monopole and dipole validation cases are provided in the Tutorials folder of the
OpenCFD-FWH project on GitHub. One can change the parameters in these programs to validate our code and get a better
understanding of OpenCFD-FWH.

\subsection{\label{sec:level4A}30P30N far-field noise prediction}
The 30P30N configuration was developed by McDonnell Douglas (now Boeing) in the early 1990s. It has been extensively used in the
study regarding the aeroacoustic characteristics of high-lift devices, especially for slat noise \cite{murayama_experimental_2014,
terracol_aeroacoustic_2015,souza_effect_2015,choudhari_assessment_2015,terracol_investigation_2016,murayama_experimental_2018,
souza2019dynamics,himeno_spod_2021}.

\begin{figure}
  \includegraphics[width=0.35\textwidth]{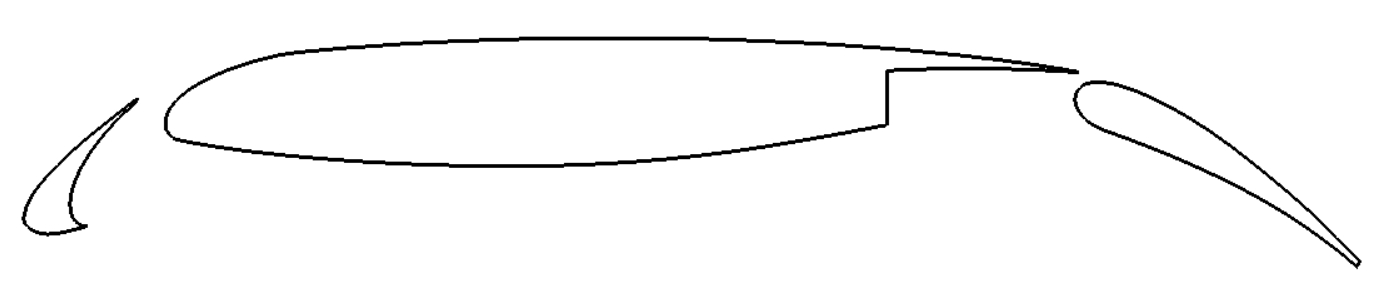}
  \caption{\label{fig:30P30N} Profile of the JAXA modified 30P30N airfoil.}
\end{figure}

The JAXA modified version 30P30N\cite{murayama_experimental_2014, murayama_experimental_2018} is utilized here to validate the code.
The airfoil profile of the 30P30N configuration is shown in Fig.~\ref{fig:30P30N}, with a stowed chord length of $c_s=0.4572\;m$.
Both the deflection angles of the slat and flap are $30^\circ$, with the chord lengths of the slat and flap being $0.15c_s$ and
$0.3c_s$, respectively.

IDDES based on SA turbulence model is carried out on the OpenCFD-EC solver. The inflow Mach number is $0.17$, with an AoA of $5.5^\circ$. 
The Reynolds number based on the stowed chord length is $1.71\times 10^6$.

\begin{figure}
  \includegraphics[width=0.42\textwidth]{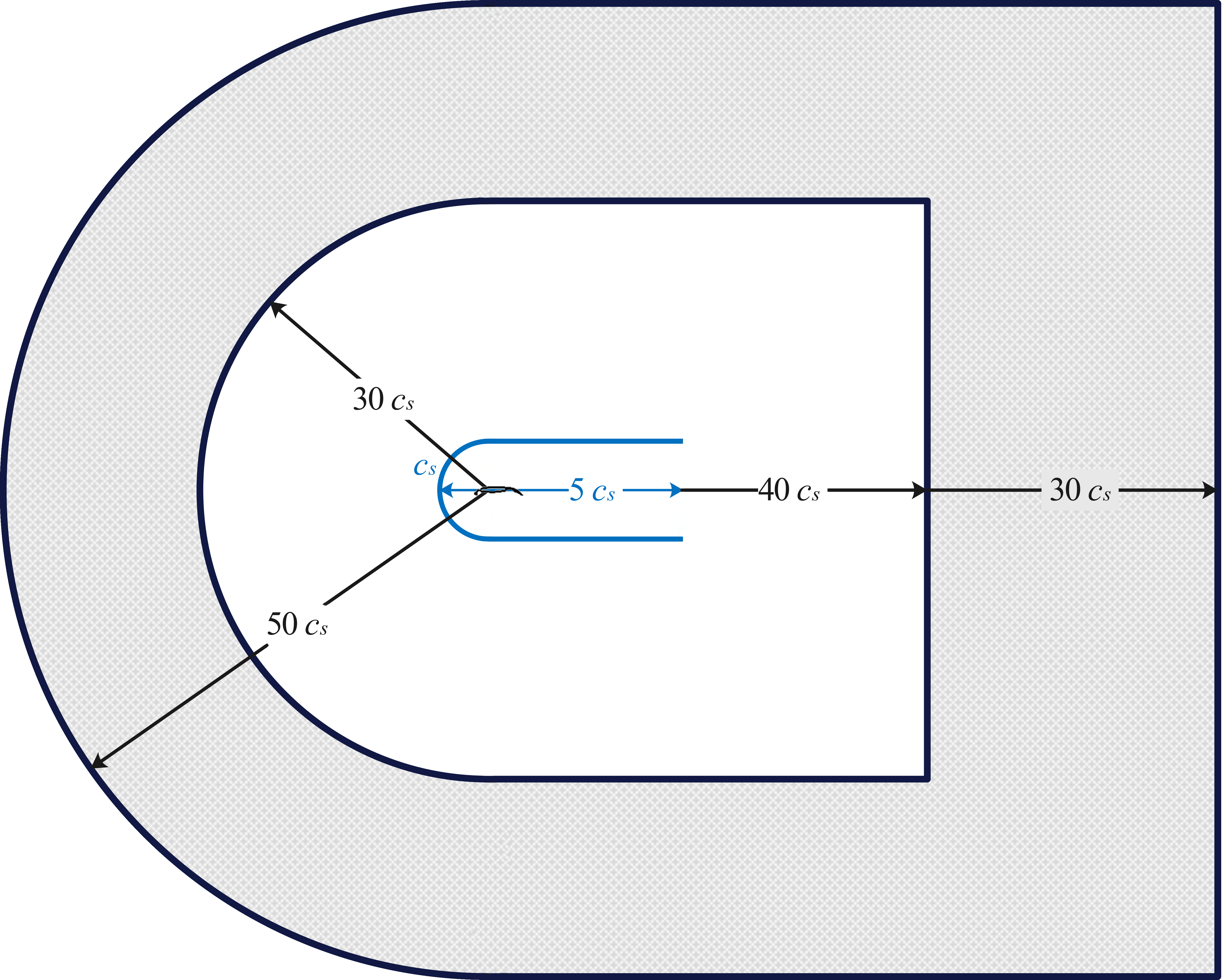}
  \caption{\label{fig:computational_domain} Spanwise cross-section of the computational domain (not in scale). 
  The grey area and blue line denote the sponge layer and the permeable FW-H data surface, respectively.
  }
\end{figure}

Fig.~\ref{fig:computational_domain} depicts a sketch of the computational domain. It extends $50c_s$ in the 
forward and vertical directions and $75c_s$ in the rear direction. Its length in the spanwise direction equals to
$1/9c_s$, following the recommendation in the BANC-III workshop\cite{choudhari_assessment_2015}
(the 3rd AIAA Workshop on Benchmark Problems for Airframe Noise Computations). A periodic boundary condition is applied
in the spanwise direction. The permeable FW-H data surface is indicated by the blue line in Fig.~\ref{fig:computational_domain},
which is one stowed chord length away from the 30P30N airfoil and stretches $5c_s$ in the wake flow direction. No endcap is
used to avoid the spurious (numerical) noise created by wake flows crossing the permeable FW-H data surface
\cite{ribeiro2023lessons}. The spanwise length of the FW-H surface is identical to the computational domain. Besides, a sponge
layer is deployed at the boundaries of the computational domain, where the viscosity is adjusted to 100 times the value used in
the physical domain to mitigate reflections at the domain boundaries, in accordance with the approach taken by Himeno et al.
\cite{himeno_spod_2021}.

\begin{figure}
  \centering
  \subfigure[]{
  \includegraphics[width=0.45\textwidth]{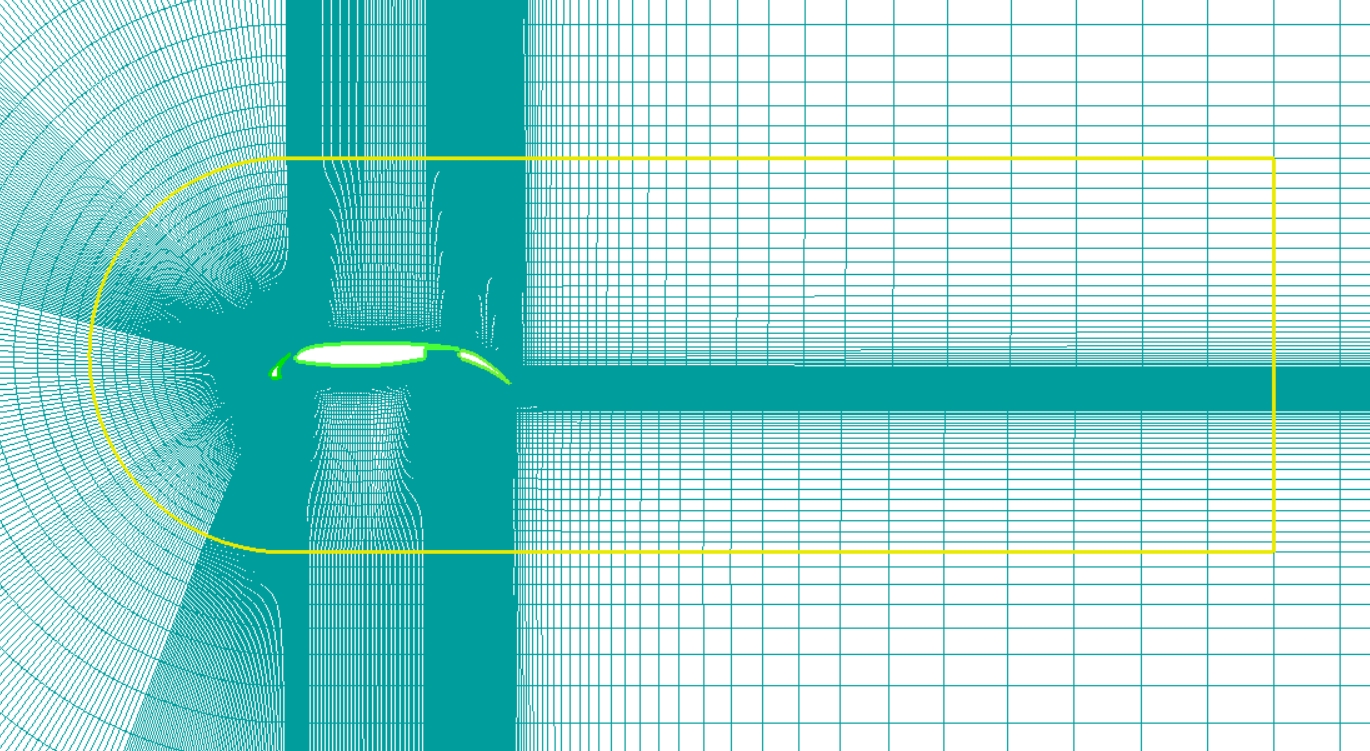}
  }
  \quad
  \subfigure[]{
    \includegraphics[width=0.45\textwidth]{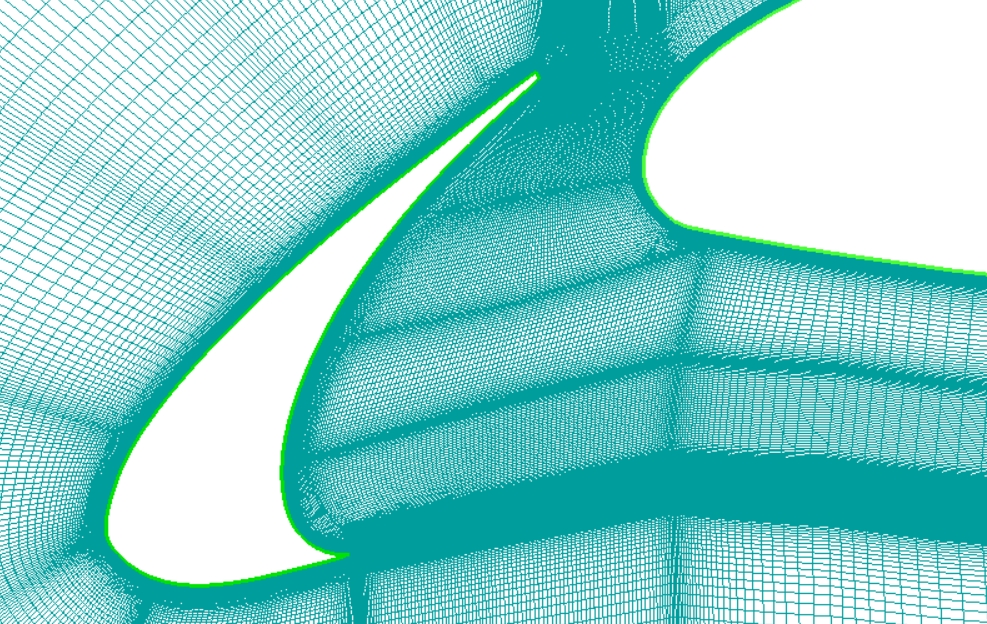}
    }
  \caption{\label{fig:mesh} Mesh details of (a) permeable FW-H data surface, (b) slat cove area.
  }
\end{figure}

A multiblock structure mesh with C-type topological is created, yielding a total cell count exceeding 43 million. Each
plane of the 2.5D mesh comprises approximately 0.25 million cells, and the entire mesh is composed of 175 planes with
equal spacing in the spanwise direction. Close-up views of the mesh around FW-H surface and slat cove area are shown in
Fig.~\ref{fig:mesh}. Additionally, the average value of the dimensionless wall distance $y+$ of the mesh is below unity.

The well-known Roe scheme is employed to decompose the inviscid flux, with the third-order MUSCL scheme for variable
reconstruction. The implicit dual-time LU-SGS method is applied for time advancement, with a time step of $2\times 10^{-7}s$.
Five inner subiterations are used, with local time-stepping approach to accelerate the convergency process. And the FW-H data
sampling interval is $1\times 10^{-5}s$. An RANS simulation with SA turbulence model is carried out to initialize the flow field.
Subsequently, approximately $0.1s$ of physical time is calculated by IDDES-SA, with $0.06534s$ available for data processing
after removing the initial transient. 

\begin{figure}
  \centering
  \subfigure[]{
  \includegraphics[width=0.46\textwidth]{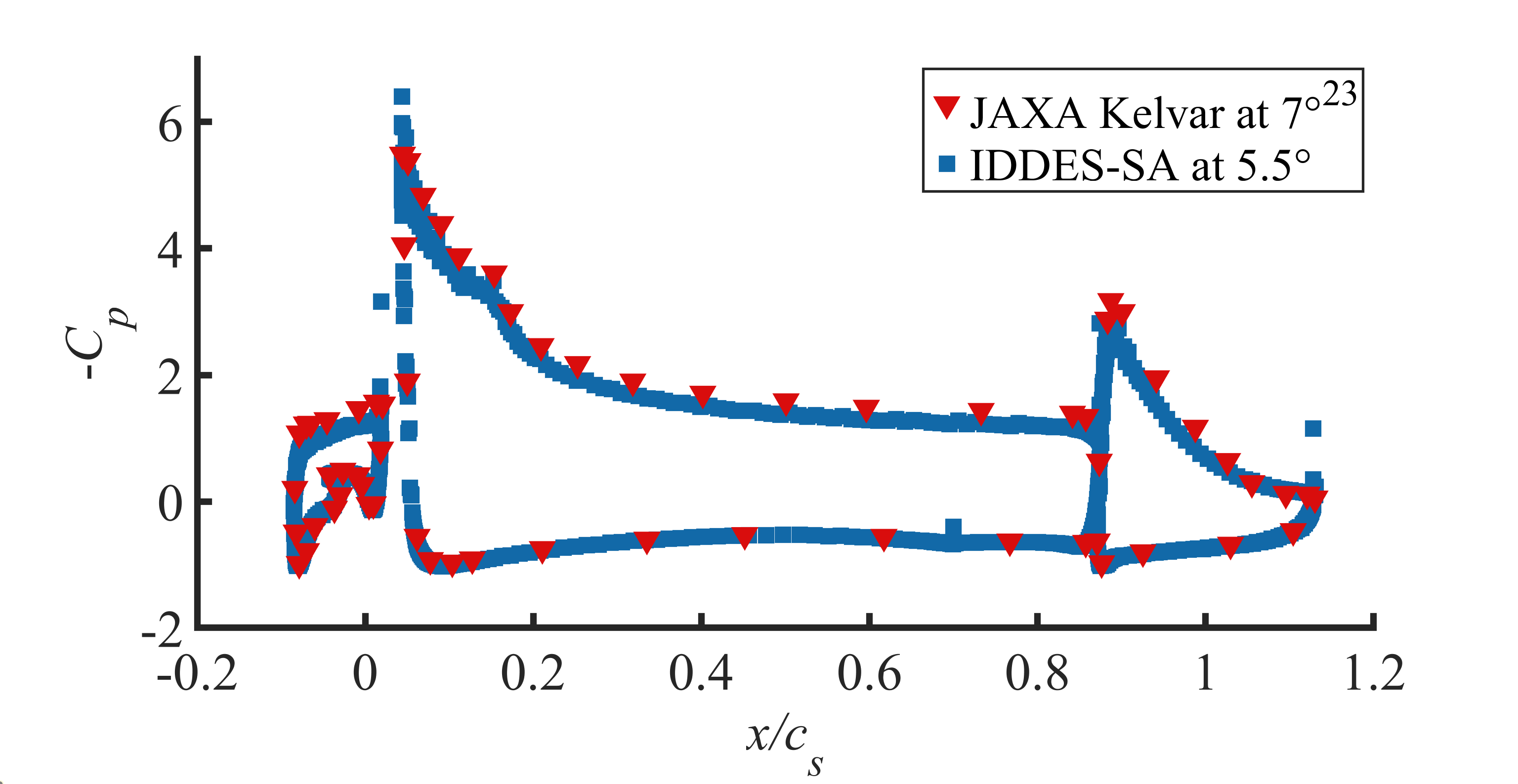}
  }
  \quad
  \subfigure[]{
    \includegraphics[width=0.46\textwidth]{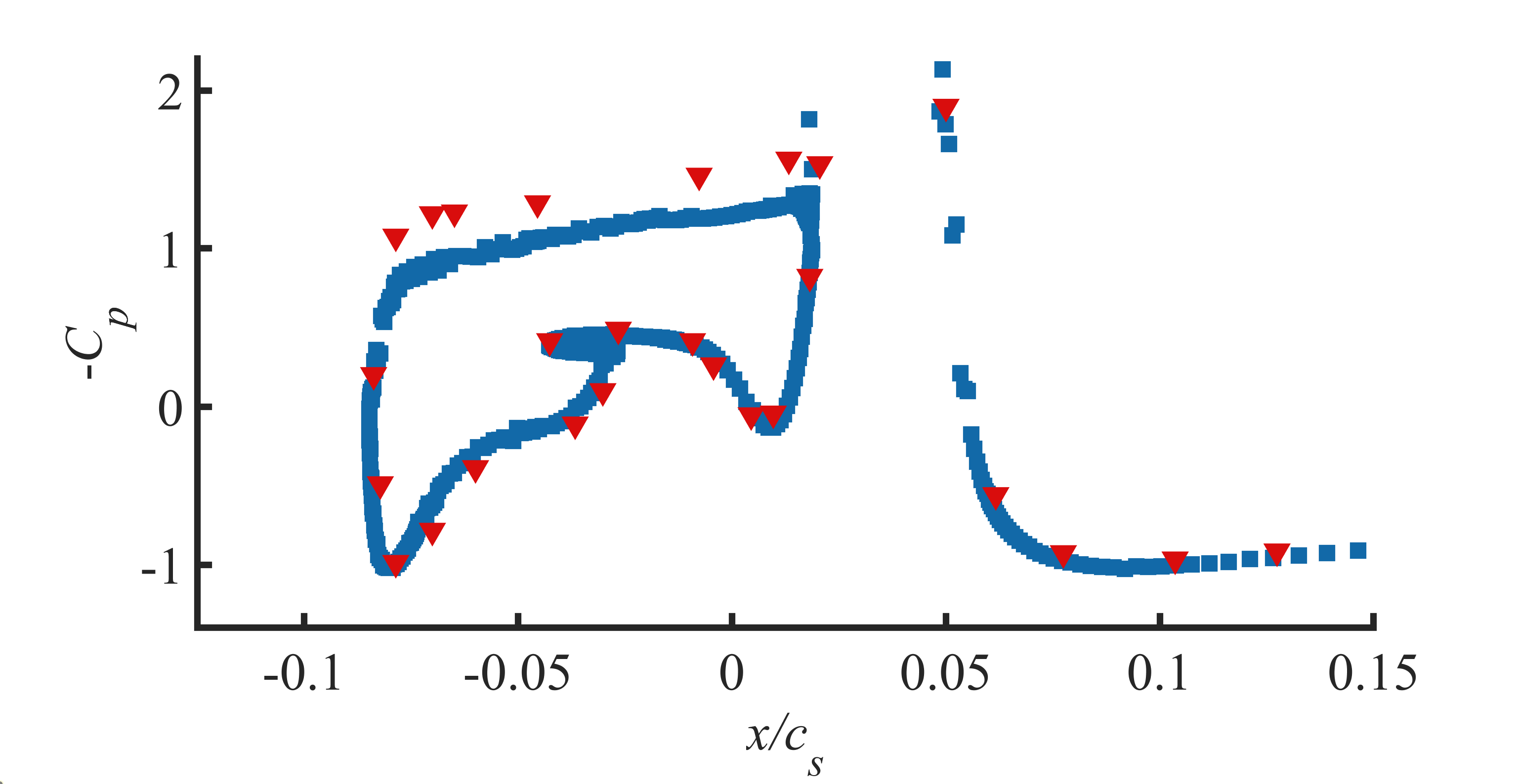}
    }
  \caption{\label{fig:Cp} Time average $C_p$ distribution at the last $0.036s$ of (a) all elements, (b) around slat.
  }
\end{figure}

A time average $C_L$ of $2.6214$ is obtained, with a difference less than $2.3\%$ compared to the average outcomes of the BANC-III 
\cite{choudhari_assessment_2015} $2.6821$. Fig.~\ref{fig:Cp} presents the time average $C_p$ distribution obtained over the last
$0.036$ seconds. While there is a slight underprediction of negative pressure on the suction side, a reasonably good agreement can
be seen, especially in the slat cove region, when compared with the JAXA Kevlar experiment\cite{murayama_experimental_2018} under
$7^\circ$ AoA. 

The scaling method used by Avallone et al.\cite{avallone2017benefits} is utilized to take into account the difference between
the spanwise acoustic integration size of the numerical simulation and experiment:
\begin{align}
  PSD_{corr}=PSD+10log_{10}\left(\frac{b_{exp}}{b_m}\right),\label{eq:PSD_corr}
\end{align}
where $b_{exp}$ and $b_m$ are equal to $650\;mm$ and $50.8\;mm$, respectively.

\begin{figure}
  \includegraphics[width=0.48\textwidth]{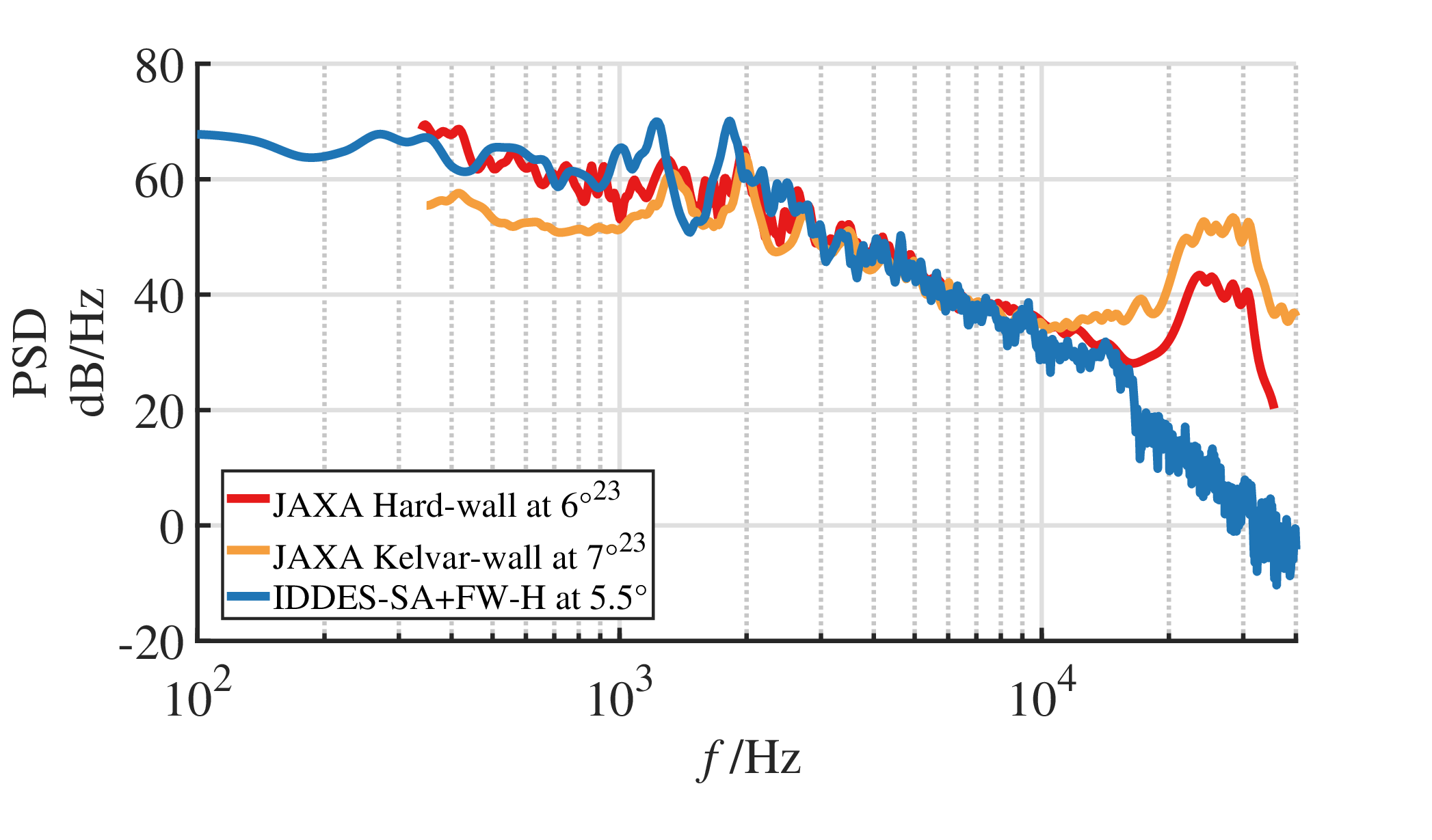}
  \caption{\label{fig:PSD} PSD of far-field acoustic signals at the center of the JAXA phased-microphone array
  \cite{murayama_experimental_2018}.
  }
\end{figure}

The pressure signal obtained by the OpenCFD-FWH code is segmented into blocks with a $50\%$ overlap, and
a Hanning window is employed. A total of 6535 sampling frames are input to the code, and the running time with different 
parallel strategies can be found in Table~\ref{tab:parallelization}. As shown in Fig.~\ref{fig:PSD}, the PSD 
(Power Spectral Density) result of the code is in good consistency with the JAXA hard-wall experiment at frequencies
below $10\;kHz$. Both the results exhibit a slightly higher noise level in the low-frequency range compared to the JAXA
kevlar-wall experiment. Furthermore, the humping noise originating from the high-frequency vortex shedding from the slat
TE of the reduced-scale wind tunnel model is absent in the FW-H result. This is attributed to the relatively coarse mesh
at the slat TE, which is unable to capture the high-frequency vortex shedding. Overall, the result validates that the
far-field noise can be accurately evaluated by the OpenCFD-FWH code.

\section{\label{sec:level5}CONCLUSIONS}
This paper presents the methodology, parallel implementation and validation of a post-processing code: OpenCFD-FWH,
designed specifically for predicting far-field noise in wind tunnel cases, encompassing nearly all scenarios
encountered in aircraft CFD cases. It is developed to use the flow field results of our OpenCFD-EC solver as input.
However, it can be readily deployed for use with other solvers by modifying the data reading part of the code or
converting the FW-H dataset to the required format. Moreover, the deployment of the code only required an MPI library
and a Fortran 90 compilation environment, without the need to install OpenCFD-EC or other affiliated libraries.

The code is based on the integration formulation of a nondimensional FW-H equation for permeable surface with convective 
and AoA effects corrected by Garrick Triangle, and 2D plane coordinate transformation, respectively. This formulation will
increase the computational efficiency compared to the original one. Additionally, the nondimensionalization of the FW-H
equation is the same as the nondimensionalization of the Navier-Stokes equations in the OpenCFD-EC solver.

MPI-OpenMP mixing parallelization is implemented to accelerate the post-processing process and reduce memory usage on a single
node/computer when deploying the code on distributed computing systems. When dealing with very large datasets, as is common in
aeroacoustic noise research related to landing gear or high-lift devices with LES, it can avoid an out-of-memory situation.
On the CAS SunRising platform, by utilizing 31 nodes, each with 32 OpenMP threads, the computing time of the code is 538.5
times faster compared to the serial implementation. This demonstrated the high operational efficiency of OpenCFD-FWH.

Three validation cases are considered in this paper. The monopole and dipole cases are compared with exact analytical solutions,
and excellent agreements are achieved. The 30P30N configuration is used in the third case, with the flow field variable produced
by IDDES-SA simulation via the OpenCFD-EC solver as input. For frequencies below $10\;kHz$, the far-field PSD result demonstrates
relatively good agreement with JAXA experiments, particularly with the JAXA hard-wall experiment. However, the result of the
code does not present the high-frequency hump observed in the experiments. This is due to the inability of the coarse mesh to
capture the high-frequency vortex shedding at the slat TE. Overall, the code is deemed validated.

The code is openly accessible on GitHub, along with the Matlab codes for the monopole and dipole validation cases to
facilitate its utilization by readers.
\begin{acknowledgments}
  This work was supported by the National Natural Science Foundation of China (Grant No. 12272024),
  National Key Research and Development Program of China (Grant Nos. 2020YFA0711800, 2019YFA0405302, 2019YFA0405300)
  and NSFC Projects (Grant Nos. 12072349, 12232018, 12202457), National Numerical Windtunnel Project, Science Challenge Project 
  (Grant No. TZ2016001), and Strategic Priority Research Program of Chinese Academy of Sciences (Grant Nos. XDCO1000000 and XDB0500301).
  
  The authors thank CNIC (Computer Network Information Center), CAS (Chinese Academy of Sciences) for providing computer time.
\end{acknowledgments}

\section*{AUTHOR DECLARATIONS}
\subsection*{Conflict of Interest}
The authors have no conflicts to disclose.

\subsection*{Author Contributions}
\textbf{Keli Zhang:} Conceptualization (lead); Data curation (lead); Investigation (lead); Methodology (lead);
Coding (lead); Writing - original draft (lead).
\textbf{Changping Yu:} Funding acquisition (equal); Supervision (lead); Writing - original draft (equal); Writing - review \& editing (equal).
\textbf{Peiqing Liu:} Funding acquisition (equal); Supervision (equal); Writing - review \& editing (equal).
\textbf{Xinliang Li:} Funding acquisition (equal); Supervision (supporting); Coding (supporting); Writing - review \& editing (equal).

\section*{Data Availability Statement}
The data that support the findings of this study are available from the corresponding author upon reasonable request.

\appendix
\setcounter{figure}{0}

\section{\label{app:A}File structures for OpenCFD-FWH}
\subsection{\label{app:A1}An example of the control.fwh file}
\begin{figure}[H]
  \centering
  \renewcommand{\thefigure}{A\arabic{figure}}
  \includegraphics[width=0.33\textwidth]{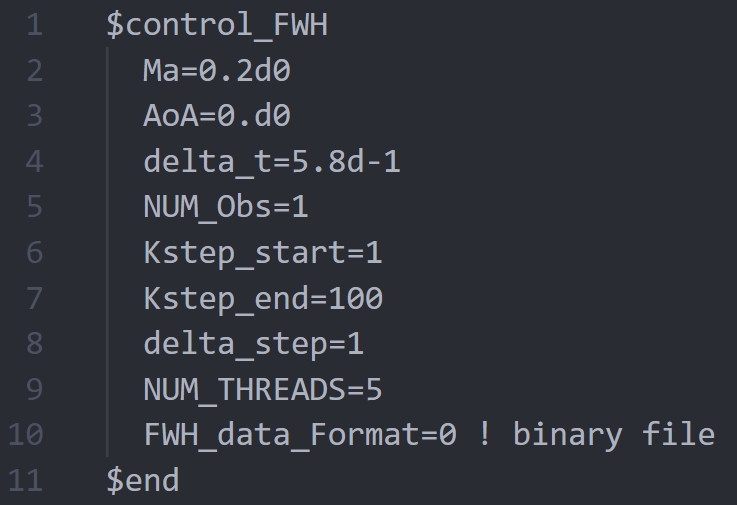}
  \caption{\label{fig:control_fwh} An example of the control.fwh file.}
\end{figure}

OpenCFD-FWH utilized namelist method to read in control.fwh file. An example of the file is presented in Fig.
\ref{fig:control_fwh}, with the values therein representing the default settings in OpenCFD-FWH. Kstep\textunderscore
start and Kstep\textunderscore end determine the start and final steps for FW-H post-processing, respectively.
With the step interval: delta\textunderscore step, OpenCFD-FWH can calculate the number of sampling frames. FWH
\textunderscore data \textunderscore Format decided whether the FW-H dataset is in binary or ASCII format
(0 for binary and 1 for ASCII).

\subsection{\label{app:A2}Structure of the FWH\textunderscore Surface\textunderscore Geo.dat file}
\begin{figure}[H]
  \centering
  \renewcommand{\thefigure}{A\arabic{figure}}
  \includegraphics[width=0.28\textwidth]{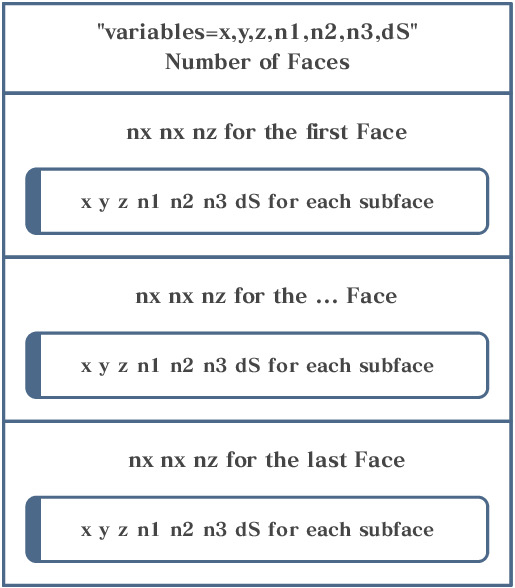}
  \caption{\label{fig:FWH_Surface_Geo} A schematic of the FWH\textunderscore Surface\textunderscore Geo.dat file. "" denotes a string of text.}
\end{figure}

FWH\textunderscore Surface\textunderscore Geo.dat file is in ASCII format in convenient for data checking. It follows
a structure similar to the Generic boundary description .inp file. It starts with a first line of text: variables=x,y,z,n1,n2,n3,dS,
as illustrated in Fig.~\ref{fig:FWH_Surface_Geo}. The second line contains a single number indicating the total number of
Faces. Subsequently, is the nx, ny, nz for one Face, along with the x, y, z, n1, n2, n3, dS values for its subfaces, until the last Face.
Note that each Face is described in a block way, consequently one of the nx, ny, nz will be 1.

\subsection{\label{app:A3}Structure of the FW-H dataset}
\begin{figure}[H]
  \centering
  \renewcommand{\thefigure}{A\arabic{figure}}
  \includegraphics[width=0.42\textwidth]{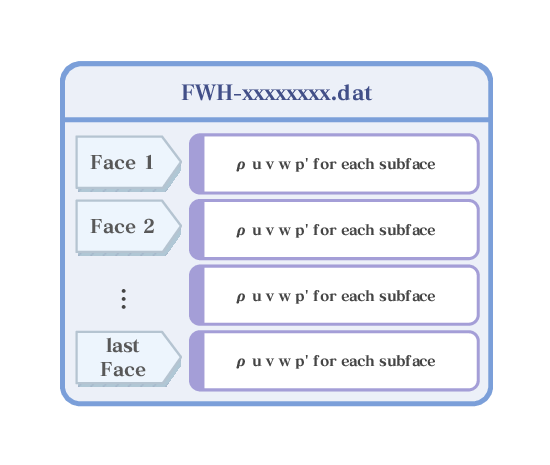}
  \caption{\label{fig:FWH-xxxxxxxx} A schematic of the FWH-xxxxxxxx.dat file.}
\end{figure}

The FW-H dataset used for OpenCFD-FWH comprised multiple FWH-xxxxxxxx.dat files, where xxxxxxxx denotes the iteration steps.
The file can be in either binary or ASCII format, with binary format being recommended for its significantly smaller data size.
A schematic of the file structure is provided in Fig.~\ref{fig:FWH-xxxxxxxx}.

\section*{REFERENCES}
\nocite{*}
\bibliography{aipsamp}

\end{document}